\documentclass{article}
\usepackage{arxiv}
\usepackage{rotating}
\usepackage{tabularx}
\usepackage{caption}
\usepackage{ragged2e}
\usepackage{longtable}
\usepackage{amsmath}
\usepackage{graphicx}
\usepackage{MnSymbol}
\usepackage{doi}
\usepackage{multirow}

\title{Comprehensive Review on Semantic Information Retrieval and Ontology Engineering}
\author{ \href{https://orcid.org/0000-0001-5054-8670}{\includegraphics[scale=0.06]{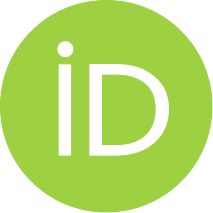}\hspace{1mm}Sumit Sharma}\thanks{The authors would like to thank National Institute of Technology Kurukshetra, India for financially supporting this research work.} \\
	Department of Computer Applications\\
	National Institute of Technology\\
	Kurukshetra, India \\
	\texttt{sumit\_6180087@nitkkr.ac.in} \\
	\And
	\href{https://orcid.org/0000-0002-7432-8506}{\includegraphics[scale=0.06]{orcid.pdf}\hspace{1mm}Sarika jain} \\
	Department of Computer Applications\\
	National Institute of Technology\\
	Kurukshetra, India \\
	\texttt{jasarika@nitkkr.ac.in} }

\date{}

\begin{document}
\maketitle
\begin{abstract}
 Situation awareness is a crucial cognitive skill that enables individuals to perceive, comprehend, and project the current state of their environment accurately. It involves being conscious of relevant information, understanding its meaning, and using that understanding to make well-informed decisions. Awareness systems often need to integrate new knowledge and adapt to changing environments. Ontology reasoning facilitates knowledge integration and evolution, allowing for seamless updates and expansions of the ontology. With the consideration of above, we are providing a quick review on semantc information retrieval and ontology engineering to understand the emerging challenges and future research. In the review we have found that the ontology reasoning addresses the limitations of traditional systems by providing a formal, flexible, and scalable framework for knowledge representation, reasoning, and inference.
\end{abstract}

\section{Introduction}

    Information retrieval (IR) is the process of obtaining relevant information from a collection of information resources either structure or unstructured~\cite{a1,singh2023privacy}. It involves searching for and retrieving information from various sources such as databases, search engines, and digital libraries~\cite{a2}. The goal of IR is to find the most relevant information based on a user's query or search terms. IR techniques include keyword-based searching, natural language processing, machine learning algorithms, reasoning and so on \cite{sharma2021lsmatch,bhargava2023ethnolinguistic}. It is used in various fields such as web search, digital libraries, e-commerce, scientific research and situation awareness~\cite{a3} etc. Information retrieval can save a lot of time by quickly providing relevant information to the user. Instead of manually searching through various sources, IR algorithms can retrieve information in a matter of seconds~\cite{a11}. By automating the process of information retrieval, it increases the efficiency of the search process. This allows users to focus on other important tasks \cite{singh2023secure}. IR algorithms are designed to provide accurate and relevant information to users. This helps to eliminate irrelevant or inaccurate information, which can save time and improve decision-making. Information retrieval can help organizations manage their knowledge more effectively by providing easy access to relevant information \cite{sharma2022lsmatch}. This can improve productivity and decision-making within the organization~\cite{a3}. IR is an important tool for researchers who need to find relevant information quickly and efficiently \cite{kumar2023power}. It can help researchers to identify new research topics, find relevant literature, and discover new insights.

    \begin{figure}[!htbp]
        \centering
        \includegraphics[width=0.96\textwidth]{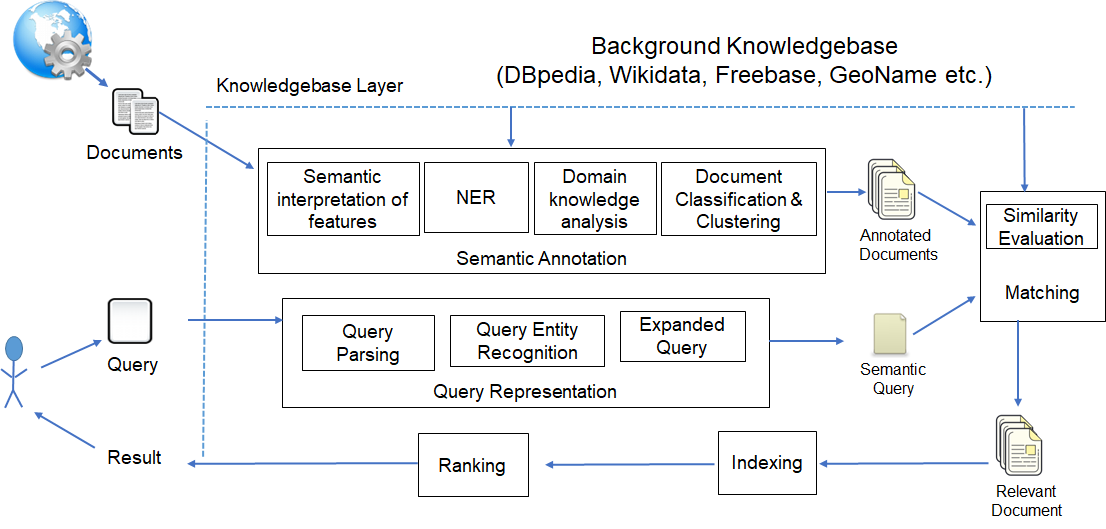}
        \caption{ Semantic Information Retrieval Architecture}
        \label{fig:fig1}
    \end{figure}

    Ontology is a formal representation of knowledge or a conceptual framework that specifies the entities, their qualities, and the interactions between them within a certain domain in the context of computer science and information science~\cite{a4,sharma2022altering,bhargava2023ethnolinguistic}. Ontologies promote IR in terms of information sharing and interoperability between different systems or applications by providing a structured and standardized manner to describe knowledge~\cite{a5}. In practical terms, ontologies are used in various domains, such as artificial intelligence (AI), semantic web (SW), information retrieval (IR), and knowledge discovery (KD). They help in organizing and structuring information, enabling better search capabilities, reasoning, and knowledge integration across different systems and applications. The information retrieval architecture depicted in Figure~\ref{fig:fig1}, explains the process of retrieving information from the semantic web. The user query is semantically mapped with the semantic information derived from the background knowledge, and the relevant information is retrieved with indexing and ranking and delivered to the user.

    \section{Emerging Challenges}
    Ontology in the context of information retrieval presents several research challenges. Some of the key challenges include
    \begin{itemize}
        \item \textbf{ C1. Information Retrieval and Knowledge Acquisition:} Acquiring and maintaining the knowledge within an ontology is a continuous challenge. The process of gathering and validating domain-specific knowledge from various sources, experts, or existing data can be time-consuming and resource-intensive. Additionally, updating and keeping the ontology up-to-date with new information or changes in the domain poses ongoing challenges.
        \item \textbf{ C2. Ambiguity:} One of the major challenges in information retrieval is dealing with ambiguity. This refers to situations where a query or search term can have multiple meanings or interpretations, making it difficult for the system to accurately retrieve relevant information.
        \item \textbf{ C3 Scalability and Adaptability:} As the amount of information available online continues to grow, IR systems must be able to scale to handle large volumes of data. This can be a challenge for systems, especially those that rely on manual indexing or classification. Ontologies need to handle large-scale and diverse information. Ensuring that ontologies can scale to accommodate a large number of entities, properties, and relationships is a challenge. Additionally, ontologies should be flexible and adaptable to accommodate changes in the domain or evolving user needs without requiring a complete reconstruction \cite{kumar2022security}.
        \item \textbf{ C4. Ontology Design and Construction:} Developing a well-designed ontology for information retrieval requires careful consideration of the domain, the target user needs, and the specific information retrieval tasks. Creating an ontology that accurately represents the concepts, relationships, and attributes relevant to the domain can be a complex task.
        \item \textbf{ C5. Semantic Search and Reasoning:} Exploiting the semantic knowledge captured in ontologies for advanced search capabilities and reasoning is an ongoing research area. Enhancing information retrieval systems to utilize ontology-based semantic search, semantic similarity, or inference techniques can improve the accuracy and relevance of retrieved results.
        \item \textbf{ C6. Heterogeneity:} Differences in the meaning, interpretation, and intended application of the same or related data, known as "Semantic Heterogeneity," result when different parties generate schema or data sets for the same domain. When building global schemas for a heterogeneous database, semantic heterogeneity between database system components is the greatest challenge.
    \end{itemize}

    Addressing these challenges we are requiring interdisciplinary research efforts involving information retrieval, knowledge representation, semantic web technologies, human-computer interaction, and domain-specific expertise. By overcoming these challenges, ontologies engineering can be effectively leveraged to improve the retrieval, organization, and understanding of information in various domains on the semantic web as described in the Figure 2.

    \begin{figure}[!htbp]
        \centering
        \includegraphics[width=0.46\textwidth]{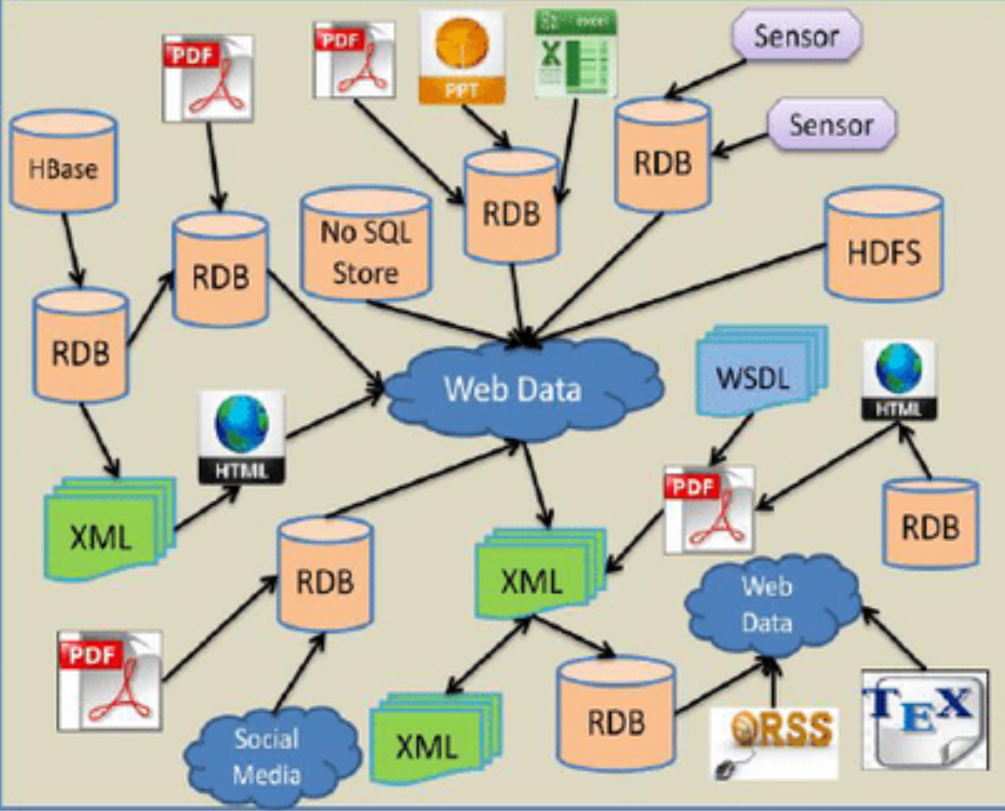}
        \caption{ Distribution of heterogeneous data on the semantic web}
        \label{fig:fig2}
    \end{figure}

    \section{Related Work} We are conducting a thorough literature review on ontology and information that involves an extensive search, analysis, and synthesis of relevant academic articles, conference papers, and scholarly publications. Various studies have proposed models that incorporate ontologies into information retrieval systems to enhance search effectiveness, precision, and recall~\cite{a6}. These models often leverage semantic knowledge captured in ontologies to improve query expansion, relevance ranking, and result clustering~\cite{a7}. Several works have focused on automatically learning or extracting ontologies from unstructured or semi-structured data sources~\cite{a8}. These approaches aim to facilitate the creation of ontologies for specific domains without manual construction. Ontologies can be used to expand user queries by adding related terms or concepts~\cite{a9}. This approach helps overcome issues like term mismatch and improves the retrieval of relevant documents by incorporating semantic relationships between concepts in the knowledge base (KB). Ontologies have been utilized to annotate and index documents with semantic metadata~\cite{a10, a11, a17, a18}. This allows for more precise document matching and retrieval based on the semantic similarity between user queries and document content~\cite{a12, a13, a19, a20,kumar2021discussion}. Researchers have investigated techniques for aligning and integrating multiple ontologies to improve interoperability and cross-domain retrieval. These methods aim to reconcile semantic heterogeneity and facilitate knowledge sharing across different ontologies~\cite{a14, a21}. Further comparative summary of related work described in Table-1.


    \captionof{table}{Comparative Summary of literature of Semantic Information Retrieval\label{tab:tb1}}

    \begin{longtable}{|p{.9cm}|p{1.5cm}|p{3.5cm}|p{2.8cm}|p{2.8cm}|p{2.5cm}|}  \hline
    Year &	Authors &	Description &	Pros &	Cons &	Future scope  \\ \hline
    
    2011 &	Bodyanskiy et.al. ~\cite{a8}	&	Word level annotate unstructured text, MPNN (Modified Probabilistic NN)	&	Reduce the dimension of search using Vector space, Fast	& Reduce the quality of data processing	&	Expend the dimension in the search space, indexing and ranking. \\ \hline

    2014 & M Schuhmacher, ~\cite{a7} & Entity ranking and computing document similarity. Use DBpedia as Ontology. Scope is Searching, ranking & Use Semantic Relation weight (WPred). Better performance, Graph Edit Distance techniques.& Take Knowledgebase as Ontology, Unable to Error stemming. & Developing an information theoretic measure to identify the semantically specific, highly informatics relation b/w entities. \\ \hline

    2016 & F Corcoglioniti, et.al, ~\cite{a9} & This method improved PIKE by simultaneously processing queries and documents. It automatically retrieved semantic content (entities, kinds, frames, temporal information) in terms-oriented and provided the scope of searching, indexing, and ranking. & Both Query and Document process, Automatically NER, Automatic indexing using semantics. & No relevance of position of the term. & There could be test on the real world environment, Use with large dataset. \\ \hline

    2016 & F Corcoglioniti ~\cite{a10} & KE framework, linguistic feature extraction and knowledge distillation are used. & Use Large database Create Single RDF Knowledge Graph. & No semantic association & Scope is Knowledge extraction, Searching. \\ \hline

    2017 & A Sayed, et.al, ~\cite{a11} & Semantic search engine support both keyword based and semantic based searching. The system is based on the RDF dataset as well as ontological graph, it supports Arabic and English. 	& Ontology based approach, Handle the heterogeneous type. & Domain Specific, use College of Applied Sciences (CAS) ontology	& Scope is Design, interference, storage, searching, indexing, query processing.\\ \hline 

    2017 & F Pech, ~\cite{a13}	& Automatically annotate unstructured text using concept similarity, TagMe tool used for NER form Wikidata. & Term level annotation. Use wiki data as ontology	Better performance & No semantic association. & Word2Vec tool for semantic extraction of term, Document level annotation. \\ \hline

    2017 & Gao, Ge et.al ~\cite{a12} & BIMTag	Based on IFC ontology, author proposed a concept based automatic semantic annotation method for online BIM (building information modeling) products. Term level and document level annotation applied. Used Latent Semantic analysis for doc level SA. & For word level sem anno used WordNet ontology and IFC ontology for document annotation. Disambiguate word sense based on local context analysis. & Irrelevant terms included in the original IFC specification that effect the performance. & Explicit semantic analysis might be applied to enhance semantic annotation and retrieval. Try to use Wikipedia as source of knowledge. \\ \hline

    2019 & F Benedetti, et.al. ~\cite{a14} & Find the inter document text similarity, based on semantic context vector. Used DBpedia, Wikidata as RDF KB. & Novel approach that find the association between the documents with any RDF knowledgebase. & Retrieval external knowledge of KB (A-box). & Term oriented and scope is Searching, Annotation and Indexing and context of keyword searching over relational Structure. (RDBMS) \\ \hline

    2019 & AA Salatino, et.al, ~\cite{a17} & (CSO) Document classifier to classify the research article by syntactic and semantically. & Unsupervised approach for automatically classifying research papers according to the (CSO). syntactic, semantic, enhancement list of matched topic. & Domain Specific and limit the domain ontology. & Expend the domain to classify the other knowledge. \\ \hline
        \end{longtable}

    The development of evaluation metrics and benchmark datasets for ontology-based information retrieval is an important area of research. These efforts focus on objectively assessing the performance and effectiveness of ontology-driven retrieval systems~\cite{a15, a16, a23}. 

    \section{Research Gap and Future Research Guidelines}
    As of my knowledge and cutoff from the Table 1, the field of information retrieval and ontology has seen significant research efforts, but there are still several research gaps that warrant further investigation. We are focus on the key research challenges in this field include: C1 (Information Retrieval and Knowledge Acquisition), C2 (Ambiguity), C3 (Scalability and Adaptability), C4 (Ontology Design and Construction), C5 (Semantic Search and Reasoning), and C6 (Heterogeneity). Based on these challenges we have summaries the literature work in the Table 2 define the used approaches and the main coverage of objective.

    \begin{table}[!htbp]
        \centering
        \caption{Analytic study of some of research work in information retrieval }
         \resizebox{\textwidth}{!}
         {\begin{tabular}{|l|c|c|c|c|c|c|c|c|c|c|}  \hline
        \multirow{2}{*}{\textbf{Author}} & \multicolumn{8}{|c|}{\textbf{Objective}} & \multicolumn{2}{|c|}{\textbf{Evaluation}} \\ \cline{2-11}
        & KB & Ontology	& Generic	& RDF	& 	Searching	& 	Indexing	& 	ranking	& 	Annotation	& 	Term		& Document\\ \hline
        
        F Benedetti \cite{a14} & $\checkmark$ & & $\checkmark$ & $\checkmark$& $\checkmark$ & &  &  &  & \\ \hline 
        
        F C \cite{a10}& $\checkmark$ & &  & & $\checkmark$ & $\checkmark$ & $\checkmark$ &  &  & $\checkmark$ \\ \hline 
        
        A Sayed \cite{a11} &  & $\checkmark$&  & & $\checkmark$ &  &  &  & $\checkmark$ & \\ \hline 
        
        F Pech \cite{a13}&  & $\checkmark$&  & &  & &  &  $\checkmark$ & $\checkmark$ & \\ \hline 
        
        AA Salatino \cite{a17}&  & $\checkmark$&  & &  & &  & $\checkmark$ &  & $\checkmark$ \\ \hline 
        
        MS \cite{a7} & $\checkmark$   &  & $\checkmark$ & &  & & $\checkmark$ & $\checkmark$ &  & \\ \hline 
        \end{tabular}}
        
        \label{tab:my_label}
    \end{table}

    While the investigating the field of information retrieval and ontology, we has seen significant advancements but there are still several research gaps that present opportunities for further exploration and development. The subsequent research guidelines are designed will fulfill the research gaps and deal with the challenges described above: 
    \subsection{G1: Handling Vague and Ambiguous Queries:} Dealing with vague and ambiguous queries is a persistent challenge in information retrieval. Ontologies can help address this issue by capturing semantic relationships and providing context for query interpretation. However, more research is required to develop robust techniques that can handle the inherent ambiguity and variability of user queries.
    \subsection{G2: Ontology-based Information Retrieval Models:} Various studies have proposed models that incorporate ontologies into information retrieval systems to enhance search effectiveness, precision, and recall. These models often leverage semantic knowledge captured in ontologies to improve query expansion, relevance ranking, and result clustering.
    \subsection{G3: Semantic Annotation and Indexing:} Ontologies have been utilized to annotate and index documents with semantic metadata. This allows for more precise document matching and retrieval based on the semantic similarity between user queries and document content.
    \subsection{G4: Ontology-Based Reasoning:} Ontologies can be used to expand user queries by adding related terms or concepts. This approach helps overcome issues like term mismatch and improves the retrieval of relevant documents by incorporating semantic relationships between concepts.
    \subsection{G5: Dynamic Ontology Evolution:} Ontologies need to adapt and evolve to reflect changes in the underlying domain. Developing techniques for dynamic ontology evolution, such as automatically updating ontologies with new concepts or refining existing relationships, is an important research area to ensure the relevancy and currency of ontologies in information retrieval systems.
    \subsection{G6: Evaluation Metrics and Benchmarking:} The development of evaluation metrics and benchmark datasets for ontology-based information retrieval is an important area of research. These efforts focus on objectively assessing the performance and effectiveness of ontology-driven retrieval systems.
        
    Addressing these research gaps will contribute to advancing the field of information retrieval and ontology, enabling more effective and efficient retrieval of relevant information in ontology engineering in various domains and application scenarios.

    \section{Research Objectives}
    Addressing these research gaps will contribute to the development of more effective, efficient, and user-friendly ontology-based information retrieval systems, enabling better access to relevant information and knowledge for users across various domains. Researchers continue to explore these areas, and advancements in the field are expected to address these challenges progressively. The main objective of this study is to enhance information retrieval systems using ontologies and address the existing research gaps. The research aims to develop novel approaches and methodologies that leverage ontological knowledge to improve the efficiency, effectiveness, and user experience of information retrieval in various domains.
    
    \subsection{RO1: Develop an Ontology Evaluation Framework:} Create a comprehensive evaluation framework to assess the performance, accuracy, and utility of ontology-driven information retrieval systems. The framework will include metrics to measure system effectiveness, user satisfaction, and retrieval efficiency that fill G6.
    \subsection{RO2: Ontology Learning and Adaptation:} Explore techniques for automatically constructing and evolving ontologies from unstructured or semi-structured data. Investigate methods to incorporate new information and adapt the ontology to changing domains to bridge the gap G4, G5.
    \subsection{RO3: Semantic annotation and indexing:} Design algorithms and data structures framework that deals with ambiguous annotations and semantic variations enable efficient processing and retrieval in large-scale and diverse datasets. The research will address scalability challenges and ensure real-time retrieval performance G1, G3.
    \subsection{RO4: Semantic Query Understanding:} Develop advanced query understanding mechanisms that utilize ontological knowledge to interpret user intent and capture deeper semantic meaning in queries. The research will focus on improving the precision and relevance of retrieved results using G2, G6.

    \section{Methodology} To achieve the research objective, the developing and implementing ontology for information retrieval is a systematic process that requires careful planning and execution. Here's an implementation plan that outlines the key steps to reach the goal represented in the Fig 3.
    \begin{figure}[!htb]
        \centering
        \includegraphics[width=.48\textwidth]{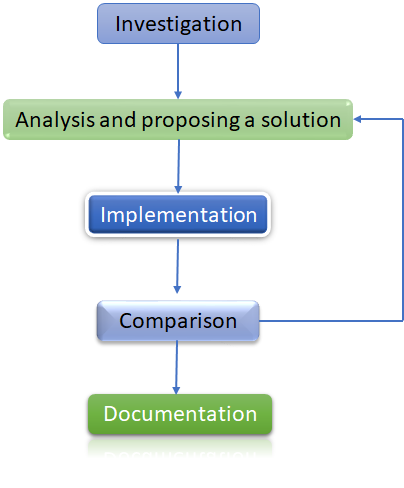}
        \caption{The step taken to accomplish the objective}
        \label{fig:fig3}
    \end{figure}
    The implementation plan should be tailored to the specific domain, organizational requirements, and available resources. Regular communication and collaboration with domain experts, users, and stakeholders throughout the development process will help ensure that the ontology and information retrieval system meet the desired objectives and effectively support information retrieval tasks.
    
    \section{Conclusions}
    Computing semantic distances in a knowledge-rich manner is the primary challenge of the web-scale ontology in providing structured representations of documents for information retrieval. This article provided a quick summary of research into generic information retrieval and ontology engineering. A comparison of these current methods is presented, and emerging challenges and potential solutions are examined. There is a need to enhance the information retrieval system to give structured representations of documents in light of the aforementioned methodologies with review and emerging challenges. In the review we have found that the ontology reasoning addresses the limitations of traditional systems by providing a formal, flexible, and scalable framework for knowledge representation, reasoning, and inference. It enhances the capabilities of awareness systems to handle complex and dynamic environments, improve interoperability, and facilitate more informed decision-making. As a result, ontology reasoning is increasingly recognized as a powerful approach for achieving effective situation awareness across various domains.

\bibliographystyle{unsrt}
\bibliography{sample} 
\end{document}